\documentclass[%
 reprint,
 amsmath,amssymb,
 aps,
]{revtex4-2}

\usepackage{graphicx}
\usepackage{dcolumn}
\usepackage{bm}
\usepackage{braket}
\usepackage[english]{babel}
\usepackage{xcolor}
\usepackage[utf8]{inputenc}
\usepackage[T1]{fontenc}

\begin{document}

\preprint{APS/123-QED}

\title{High-Tc Superconductivity in Functionalized Out-of-Plane Ordered Double Transition Metal MXenes}

\author{Mohammad Keivanloo$^{1}$}
\author{Fateme Dinmohammad$^{1}$}
\author{Shashi B. Mishra$^{2}$}
\author{Mohammad Sandoghchi$^{1}$}
\author{Mohammad Javad Arshia$^{1}$}
\author{Mitsuaki Kawamura$^{3}$}
\author{Elena R. Margine$^{2}$}
\author{Muhammad Haris Mahyuddin$^{4}$}
\author{Hannes Raebiger$^{3}$}
\author{Reza Pamungkas Putra Sukanli$^{5}$}
\author{Kenta Hongo$^{5}$}
\author{Ryo Maezono$^{6}$}
\author{Mohammad Khazaei$^{1,}$$^{7}$*}

\affiliation{}

\affiliation{$^{1}$Department of Physics, University of Tehran, North Kargar Ave, Tehran 14395547, Iran}
\affiliation{$^{2}$Department of Physics, Applied Physics and Astronomy, Binghamton University-SUNY, Binghamton, New York 13902, USA}
\affiliation{$^{3}$Yokohama National University, Department of Physics, 240-8501 Yokohama, Japan}
\affiliation{$^{4}$Quantum and Nano Technology Research Group, Faculty of Industrial Technology, Institut Teknologi Bandung, Jl. Ganesha 10 Bandung 40132, Indonesia}
\affiliation{$^{5}$Research Center for Advanced Computing Infrastructure, JAIST, Nomi, Ishikawa, 923-1292, Japan}
\affiliation{$^{6}$Institute of Science Tokyo, Graduate Major in Materials and Information Sciences, 2-12-1-S6-22 Ookayama, Meguro-ku, Tokyo 152-8550, Japan}
\affiliation{$^{7}$School of Quantum Physics and Matter, Institute for Research in Fundamental Sciences (IPM), Tehran 19538-335, Iran}
\affiliation{Email: \href{mailto:mohammad.khazaei@ut.ac.ir}{\color{blue}mohammad.khazaei@ut.ac.ir}}


\date{\today}

\begin{abstract}
Two-dimensional (2D) superconductors attracted growing interest in condensed-matter physics research. In this work, we explore the superconducting properties of surface-functionalized, out-of-plane ordered double transition-metal MXenes (o-MXenes), which exhibit distinctive structural and electronic characteristics. Using first-principles calculations, we investigate the effects of electronic structure, electron–phonon coupling (EPC), anharmonicity, and anisotropy effect in superconductivity properties of o-MXenes. We examine a wide range of o-MXene systems, M$_{2}$M$^\prime$X$_{2}$T$_{2}$ (M = Mo, W; M$^\prime$ = Sc, Ti, V, Mo, Zr, Nb, Ta; X = C, N), functionalized with F, O, Cl, and H groups. Out of 128 candidates, 32 compounds are found to be mechanically, dynamically, and thermodynamically stable, exhibiting superconducting transition temperatures (T$_{c}$) from 0.1 K to 52 K. Notably, the Mo$_{2}$ScN$_{2}$O$_{2}$ compound achieves the highest T$_{c}$ of 52 K, with a superconducting gap of $\sim$10 meV. Solving the anisotropic Eliashberg equation reveals that Mo$_{2}$ScN$_{2}$O$_{2}$ is an anisotropic two-gap superconductor, and incorporating anharmonic effects decreases its T$_{c}$ slightly. We further analyze flat-band-induced EPC enhancement and present EPC matrix elements as functions of phonon wavevector q for distinct vibrational modes that show anharmonic behavior of these materials.
\end{abstract}

\maketitle


\section{Introduction}\label{sec1}
Two-dimensional materials have attracted considerable attention owing to their unique electronic and phononic characteristics.\cite{PhysRevB.111.094518}\cite{doi:10.1126/science.aab2277} Additionally, the emergence of 2D superconductors has sparked enthusiasm for innovative superconductors, with critical temperatures (T$_{c}$) potentially reaching as high as 100~K.\cite{sbbm-hvc4}\cite{doi:10.1021/acsami.3c10564} The discovery of various compounds and the evaluation of their T$_{c}$ have been conducted through theoretical methods, particularly through first-principles calculations, which facilitate accurate assessments of crystal structures, electronic configurations, and electron-phonon interactions, thus pinpointing promising superconducting materials within BCS theory for future synthesis.\cite{tian2024few}\cite{D5MH00177C}\cite{D4NR04231J}

One of the newly developed 2D materials, MXenes, offers a wide range of potential applications in electronic devices\cite{LI2021377}, ion batteries\cite{doi:10.1021/acsami.4c10656}\cite{SUNKARI2024112017}, supercapacitors\cite{POURJAFARABADI2024109811}\cite{JAMEHBOZORG2025117666}, electromagnetic shielding\cite{iqbal2024mxenes}, catalytic processes\cite{TEHRANI202427202}\cite{tehrani2025enhanced}, composites\cite{tehrani20255}, and more. MXenes are typically derived from their corresponding MAX phases, which are recognized as a family of layered transition metal carbides and nitrides \cite{khazaei2013novel}\cite{doi:10.1126/science.abf1581}\cite{keivanloo2024study}. In this context, M denotes the transition metal (for example, Ti, Nb, Mo, V, W, etc.), A represents the main group element in the periodic table (such as Al, Ga, Si, etc.), and X stands for C or N. The selective etching of the A element atoms from MAX phases using acids, like a weak hydrofluoric (HF) solution, results in the formation of a 2D MXene \cite{khazaei2013novel}\cite{doi:10.1021/acs.jpclett.0c03710}. Depending on the acid utilized, MXenes are terminated with a combination of H, OH, O, F, and Cl functional groups \cite{pang2019universal}\cite{li2018fluorine}. Pure MXenes primarily exhibit metallic properties. However, by selectively terminating MXenes with different surface groups, their stability can be enhanced, and their properties can be modified from metallic to semimetallic or semiconducting. Some metallic MXenes have even been identified as superconducting. While the highest theoretically predicted critical temperature for a pure MXene (Mo$_{2}$N) is approximately 16 K, experimental realizations such as Cl-terminated Nb$_{2}$C have shown superconductivity at lower temperatures, around 10 K\cite{doi:10.1126/science.aba8311}\cite{D2NR01939F}. MXenes leverage the potential for diverse covalent surface terminations, unveiling new opportunities for the rational electronic engineering of 2D MXenes, several of which demonstrate superconducting behavior\cite{doi:10.1126/science.aba8311}\cite{D0NR03875J}\cite{xu2024anisotropic}\cite{fan2025organometallic}.

In addition to the possibilities for electronic engineering via the manipulation of surface groups, the extensive range of elemental composition engineering in MAX phases promotes the creation of various 2D MXenes. To put it simply, modifying the types of transition metals can greatly influence the electronic characteristics of MXenes\cite{C7TC00140A}. Traditionally, MAX phases have employed a single type of transition metal\cite{DAHLQVIST20241}. However, recent progress has enabled the development of novel MAX phases that integrate two or more types of transition metals, leading to the formation of new 2D MXenes with varying transition metal compositions\cite{HUSSAIN2024101382}\cite{ALIBAGHERI2025102893}\cite{LI2023119001}. In this context, MAX phases that feature out-of-plane ordered double transition metals, known as o-MAX \cite{doi:10.1021/acsnano.5b03591} with the chemical formulas M$_{2}$M$^\prime$AC$_{2}$ and M$_{2}$M$^\prime$$_{2}$AC$_{3}$, as well as those with ordered in-plane double transition metals, termed i-MAX phases \cite{PhysRevMaterials.3.053609} with the chemical formula (M$_{2/3}$M$^\prime$$_{1/3}$)$_{2}$AC, along with their respective ordered 2Ds, o-MXenes and i-MXenes, have significantly enhanced the electronic diversity of the family of MAX phases and MXenes. Notably, o-MXenes and i-MXenes are derived from a conventional etching process applied to their corresponding MAX phases. It is important to highlight that high-entropy MAX phases, and consequently MXenes with a variety of transition metal options, have also been successfully synthesized in laboratory thereby showcasing the extensive experimental synthesizability of this material family\cite{tan2017high}\cite{anasori2014fabrication}. 

Traditional functionalized MXenes, defined by the chemical formula M$_{2}$CT$_{2}$, have already shown their promise as viable candidates for phonon-mediated superconductivity\cite{xu2024anisotropic}\cite{KOTMOOL2024416551}\cite{jamwal2025mixed}. This raises interest in how the superconductivity of functionalized o-MXene structures, incorporating two distinct transition metals, could be advantageous. To date, several o-MXenes have been successfully synthesized from the o-MAX phases, including Mo$_{2}$ScC$_{2}$T$_{2}$\cite{meshkian2017theoretical}, Mo$_{2}$TiC$_{2}$T$_{2}$\cite{doi:10.1021/acsnano.5b03591}, Mo$_{2}$Ti$_{2}$C$_{3}$T$_{2}$\cite{doi:10.1021/acsnano.5b03591}, Cr$_{2}$TiC$_{2}$T$_{2}$\cite{liu2014crystal}, and Cr$_{2}$VC$_{2}$T$_{2}$\cite{wyatt2023design}. However, based on computational predictions and the structural features of these synthesized o-MXenes, additional members of this material family show significant potential for successful synthesis in the future. In numerous investigations of superconductivity in diverse two-dimensional materials, including MXenes, anharmonic and anisotropic effects on T$_{c}$ have frequently been neglected. Nevertheless, recent studies indicate that anharmonic lattice dynamics and anisotropy can substantially affect essential characteristics such as electron-phonon coupling and, thereby, the superconducting properties across a wide range of materials\cite{tian2024few}\cite{hamidi2025prediction}.

In this research, we first evaluate the structural stability of over 128 o-MXenes, identifying all dynamically stable configurations. Additionally, we analyze electron-phonon coupling (EPC) from a dual perspective examining both electronic structure and phononic behavior—and their influence into superconducting properties. We have incorporated anharmonic effects using the stochastic self-consistent harmonic approximation (SSCHA), resulting in significant alterations in the phonon spectrum. Considering that low-dimensional systems with anisotropic Fermi surfaces frequently exhibit anisotropic electron-phonon coupling (EPC), which in turn leads to anisotropic superconductivity, we further investigated the directional dependence of their superconducting properties. Our findings reveal that Mo$_{2}$ScN$_{2}$O$_{2}$ achieves the highest superconducting transition temperature (T$_{c}$), reaching approximately 52 K.

\section{Results and Discussion}
\subsection{o-MXene Structure and Stability}
\begin{figure*}
\includegraphics[width=0.8\linewidth]{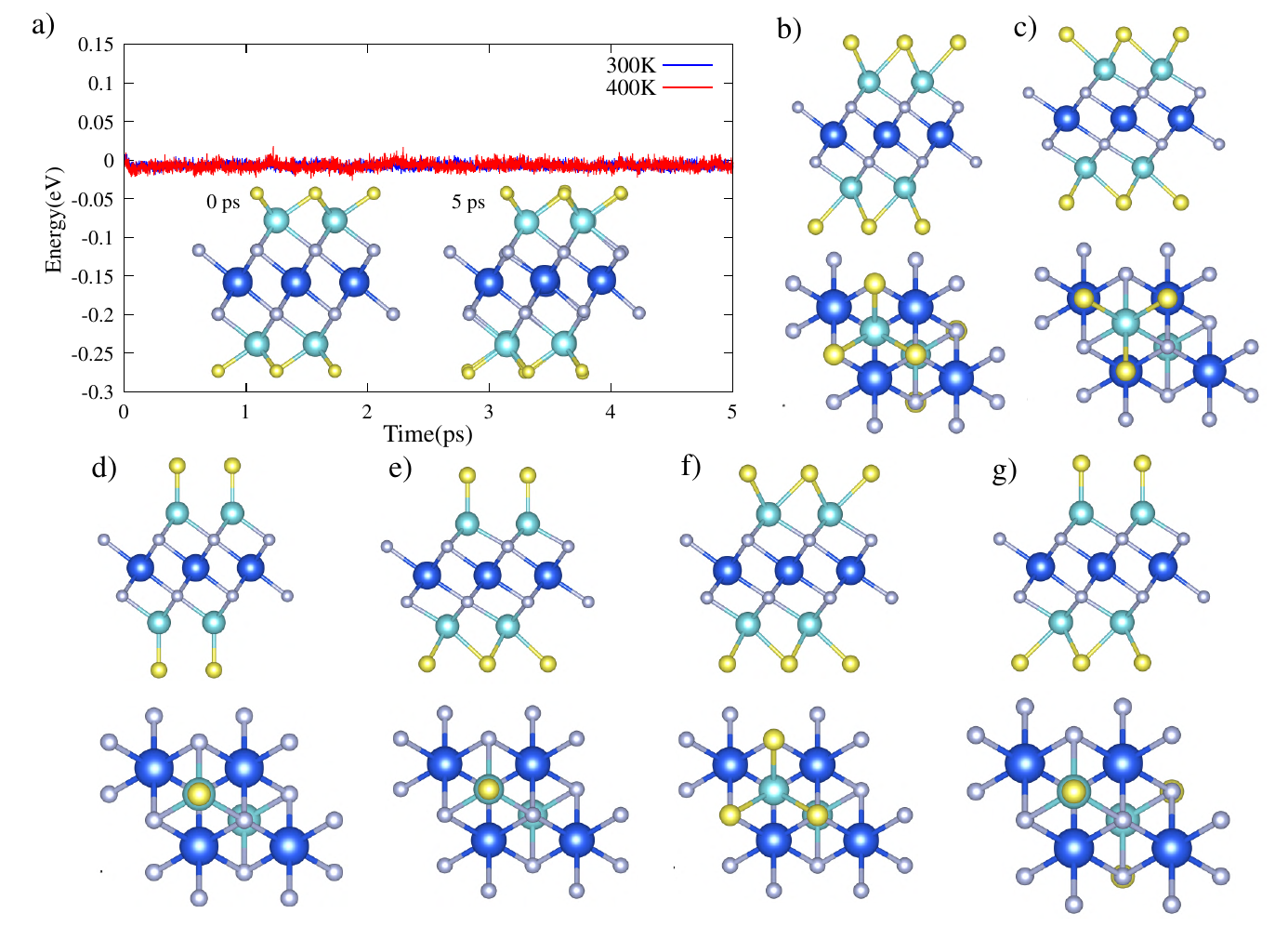}
\caption{\label{fig:wide}a) The total free energy as a function of ab initio molecular dynamics (AIMD) simulation time for a 2$\times$2$\times$1 supercell at
300K and 400K for Mo$_{2}$ScN$_{2}$O$_{2}$. The insets display corresponding structure snapshots at 0 and 5 ps, respectively.b-g) Six distinct configurations with different placements of surface chemical groups for each M$_{2}$M$^\prime$C$_{2}$ compound.}
\label{fig1:structure}
\end{figure*}

Figure \ref{fig1:structure}b-g depicts the crystal structure of o-MXene compounds, showing various positions of surface functionalization. These structures display the symmetry inherent to the hexagonal crystal system and belong to space group P-3m1~(164). As depicted in Fig.\ref{fig1:structure}, the M$_{2}$M$^\prime$C$_{2}$T$_{2}$ MXene showcases a hexagonal lattice composed of five alternating atomic layers arranged in the sequence M–C–M$^\prime$–C–M. In this configuration, the outer layers are consistently occupied by M transition metals, while the central layer is filled with M$^\prime$ transition metals. Carbon atoms are positioned in interstitial sites between the M and M$^\prime$ layers, located at the center of octahedral cages formed by the surrounding M and M$^\prime$ transition metal atoms. In this research, we examine compounds where Mo and W occupied the M sites; Sc, Ti, V, Y, Zr, Nb, Hf, and Ta (non-magnetic elements) were chosen for the M$^\prime$ sites; C and N were considered as X elements; and H, O, F, and Cl acted as surface terminations. This results in a total of 128 unique o-MXene compounds for analysis. It is well established that functionalized groups on the surfaces can adopt various configurations on the surfaces of o-MXenes according to the surface hollow sites. Consequently, in the preliminary phase of this study, we compared the total energies of six distinct configurations with different placements of surface chemical groups for each M$_{2}$M$^\prime$C$_{2}$ compound (Figure~1S). The results are presented in Table S1. Based on this analysis, structures with the lowest total energies were chosen for further investigation. Given that structural stability is fundamental to all subsequent characterizations, computational methods were utilized to thoroughly evaluate the dynamic, thermodynamic, and mechanical stability of the selected compounds.

 A total of 128 o-MXene compounds were structurally optimized at ambient pressure, followed by assessments of their dynamic stability through phonon calculations. These calculations utilized density functional perturbation theory (DFPT). Our results show that 40 of the analyzed phases do not exhibit any imaginary frequencies, confirming their dynamic stability, whereas the other 88 compounds show imaginary modes, indicating dynamic instability. According to Table 3, nitride-based o-MXene compounds that are functionalized with Cl and H do not demonstrate dynamic stability. This instability is linked to electronic interactions occurring near the Fermi level, which will be discussed in the subsequent section.

To assess the thermodynamic stability of the chosen o-MXene compounds, we conducted ab initio molecular dynamics (AIMD) simulations at 300 K for a period of 5 ps. As illustrated in Figure \ref{fig1:structure}a for Mo$_{2}$ScN$_{2}$O$_{2}$ and in Figure 3S for all other o-MXenes, the metal framework atoms display only slight oscillations around their equilibrium positions, whereas the functional groups remain firmly attached to the surface. Furthermore, the total free energy is maintained within a limited fluctuation range during the simulation, collectively affirming the thermodynamic stability of these structures. Results show that 7 of the dynamically stable phases are not thermodynamically stable, and we have 33 dynamically and thermodynamically stable o-MXene.

\begin{figure*}
\includegraphics[width=1.0\linewidth]{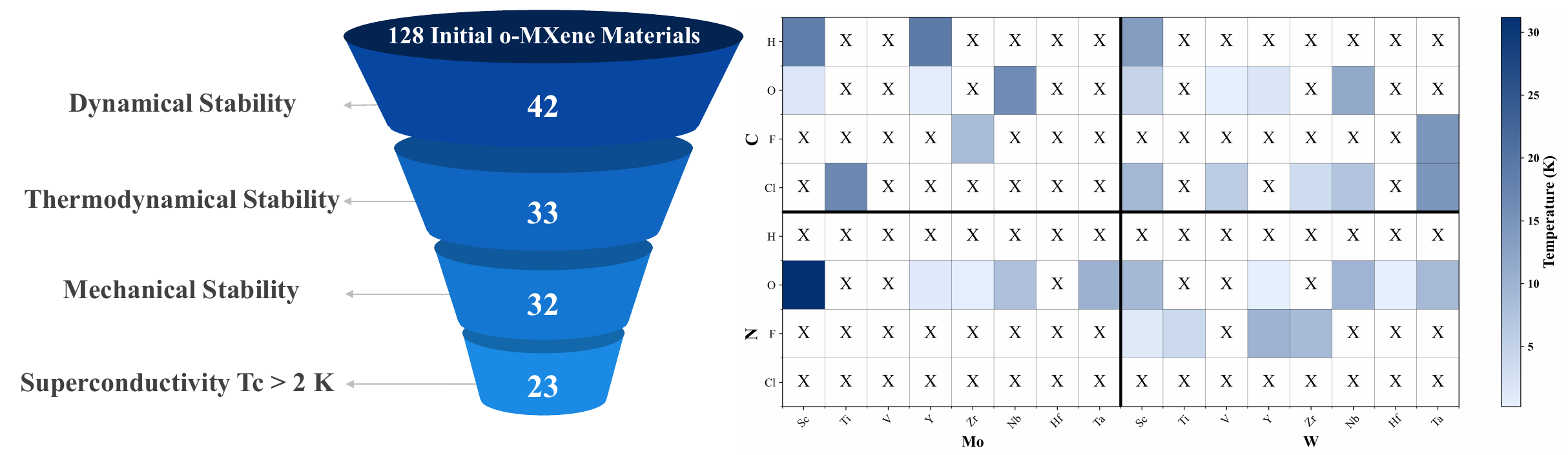}
\caption{\label{fig:wide}Stability screening of o-MXenes and a map of the superconducting critical temperature (T$_{c}$), calculated using the McMillan equation with a Coulomb repulsion parameter $\mu$=0.1, is shown for all o-MXenes. The horizontal axis represents the transition metals M and M$^\prime$, while the vertical axis corresponds to the functional groups and X atoms. Phases identified as dynamically unstable from phonon calculations are marked with $'$X$'$.}\label{fig2}
\end{figure*}

 In order to assess the mechanical stability of the studied o-MXene compounds, we computed their elastic constants, which are detailed in Table 1S. The findings show that the elastic constants meet the Born stability criteria for two-dimensional hexagonal systems\cite{PhysRevB.90.224104}, namely: C$_{11}$-C$_{12}$$>$ 0, C$_{11}$+2C$_{12}$$>$ 0, and C$_{44}$$>$ 0. These criteria verify that, except for one compound, all o-MXene structures are mechanically stable (Figure~\ref{fig2}). Furthermore, Table 2S provides a summary of the calculated formation energies for all examined o-MXene phases, which consistently show notably negative values. The close correlation between the formation energies of theoretical compounds and those that have already been synthesized indicates a strong likelihood of experimental success and considerable synthesis potential within this material family.

\subsection{Electronic Structure}
\begin{figure*}
\includegraphics[width=1.0\linewidth]{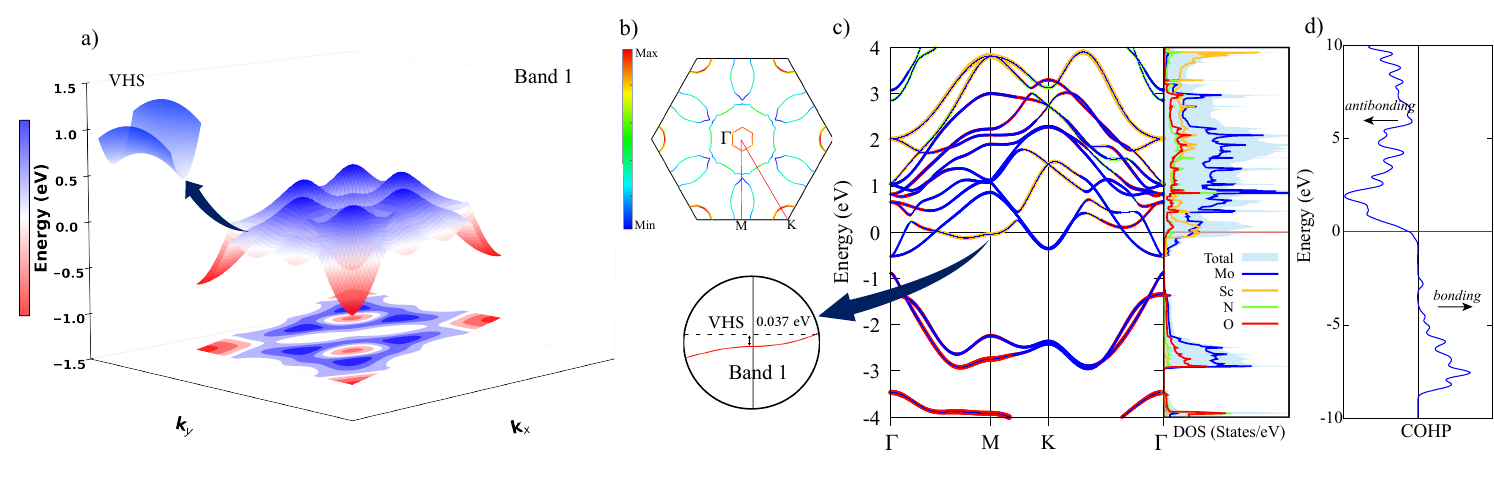}
\caption{\label{fig:wide}a) The 3D view of the flat band that has a major role in EPC and the saddle-type Van Hove singularity.b) Fermi surface, the color drawn indicates the relative Fermi velocity v$_{F}$. c) partial electronic band structure, and partial density of state of Mo$_{2}$ScN$_{2}$O$_{2}$.The Fermi energy is set to 0 eV. d) Crystal orbital hamiltonian population (COHP) of Mo$_{2}$ScN$_{2}$O$_{2}$.}\label{fig3}
\end{figure*}

To elucidate the origins of the significant electron-phonon coupling (EPC) constants and the elevated superconducting transition temperatures observed in certain o-MXene compounds, we conducted detailed calculations of their electronic and vibrational structures. The orbital-resolved band structures, partial electronic density of states (DOS), and Fermi surfaces for all studied o-MXenes are presented in 
Figures~\ref{fig3} for Mo$_{2}$ScN$_{2}$O$_{2}$ and in Figures 4S and 5S for all other stable o-MXenes. All o-MXenes exhibit metallic behavior, primarily arising from degenerate band crossings near the Fermi level, predominantly contributed by the d orbitals of the transition metals M and M$^\prime$. In these compounds, the surface-terminating groups F, O, Cl, and H-interact differently with the underlying structure due to variations in their electron affinity, and bonding configurations. These interactions lead to distinct electronic structures, as each group stabilizes its adsorption site via differential electron uptake from the surface.

The density of states at the Fermi level, N(E$_{F}$), along with the topology of the Fermi surface, are essential elements that affect superconductivity and electron-phonon interactions. In conventional MXenes, the properties of surface charge largely remain constant despite changes in layer thickness; surfaces that are terminated with F, Cl, O, or H maintain their original charge polarity regardless of thickness. Likewise, the termination groups in ordered double-transition-metal MXenes (o-MXenes) display similar charge polarity characteristics as those seen in traditional MXenes. Nevertheless, the presence of two chemically distinct transition metals in o-MXenes significantly alters the topology of the Fermi surface, introducing new electronic properties in comparison to traditional MXenes that contain only one type of transition metal. In this context, the density of states (DOS) analyses for stable o-MXenes (Figure 4S) indicate that the transition metals M and M$^\prime$ make considerable contributions to the electronic states close to the Fermi level. Surface terminations with F, O, or Cl lead to downward shifts of the Fermi level, whereas H terminations cause upward shifts when compared to o-MXenes lacking surface functionalization. These energy shifts arise from the differences in electronegativity between F, O, Cl, and H relative to the electronegativity of the surface transition metals, such that F, O, and Cl tend to attract electrons from the surface transition atoms, while H, conversely, donates electrons. The strength of the bonds between the atomic orbitals of the termination groups and transition metals, or the C orbitals and adjacent transition metals, also influences the positioning of the energy states in relation to the Fermi energy. Overall, due to the strong bonds formed between the surface transition metals (Mo/W) and Cl, O, or F, their associated hybridized states are found at lower energy levels, thus enhancing the contribution of M elements to N(E$_{F}$). Notably, in Mo$_{2}$ScN$_{2}$O$_{2}$, the comprehensive interactions of atomic orbitals position the Fermi level close to a flat band, which may increase electron-phonon interactions and improve superconducting properties.

\begin{figure*}
\includegraphics[width=1.0\linewidth]{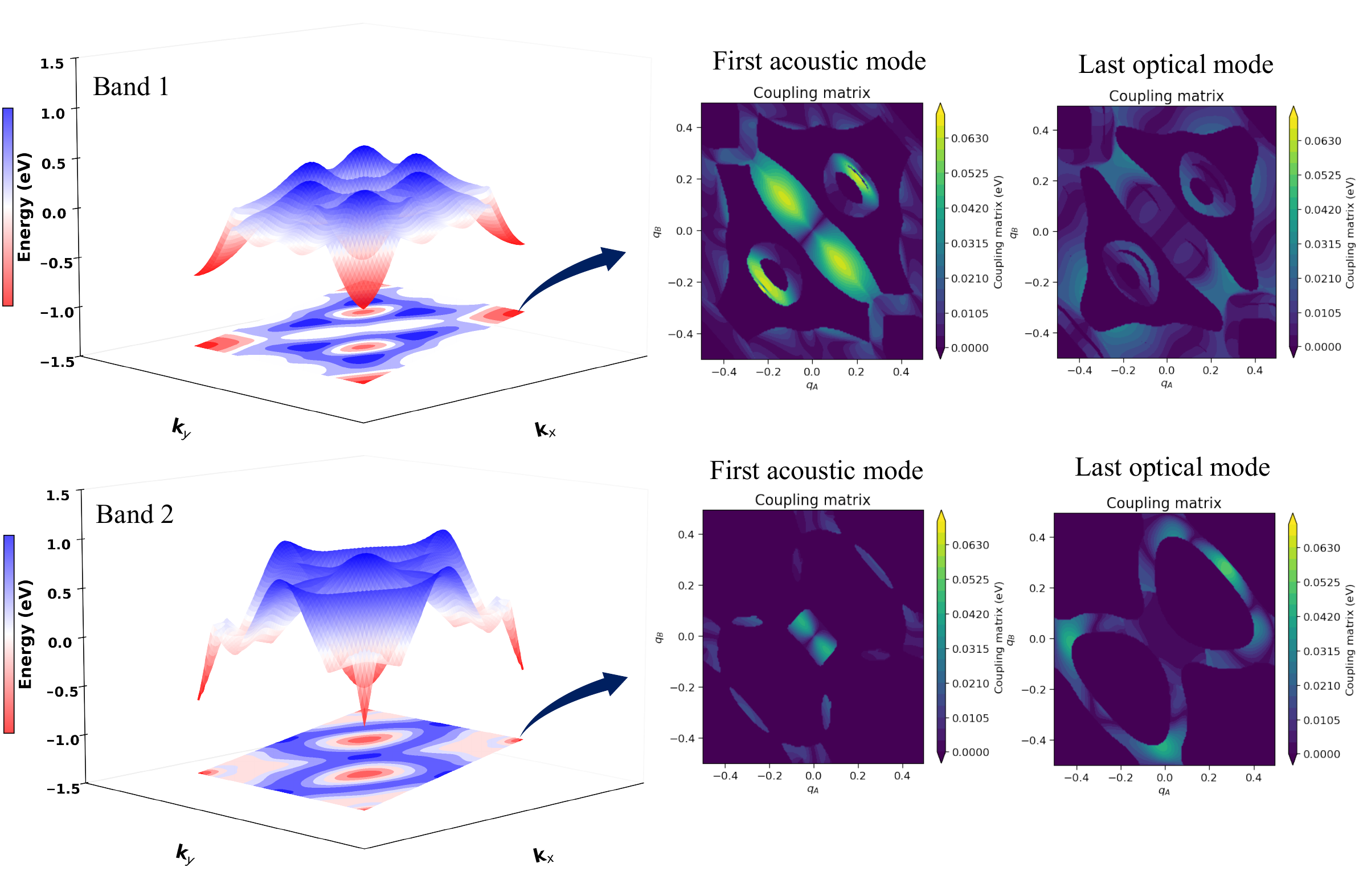}
\caption{\label{fig:wide}A 3D representation of the electronic bands crossing the Fermi level in Mo$_{2}$ScN$_{2}$O$_{2}$ and cross section of that at 0 eV, along with the corresponding electron–phonon coupling (EPC) matrix elements as functions of the phonon wavevector q, shown for representative vibrational modes—the lowest acoustic and highest optical branches.}\label{fig4}
\end{figure*}

In the case of Mo$_{2}$ScN$_{2}$O$_{2}$, which exhibits the highest predicted T$_{c}$ among the studied o-MXene compounds, the band structure features two valence bands intersecting the Fermi level. These bands, labeled as Band 1 and Band 2 (Figure~\ref{fig3}b,c), cross the Fermi surface in all $\Gamma$-M, M-K, and K-$\Gamma$ paths. At the M-point within the Brillouin Zone, a saddle-type Van Hove singularity (VHS) emerges near a half-filled band, where the Fermi surface intersects the zone boundary and electronic conduction transitions from electron-like to hole-like behavior. The VHS is located approximately below the Fermi level. In two-dimensional materials such as o-MXenes, a VHS typically consists of a saddle point linking electron- and hole-like bands. According to prior studies, when a VHS is positioned near the Fermi energy, various correlated electronic phases—including superconductivity\cite{luo2023unique}, charge density waves (CDW)\cite{PhysRevB.104.045122}, and ferromagnetism\cite{PhysRevB.92.085423}—can be significantly enhanced. Furthermore, the presence of a VHS leads to a divergence in the density of states (DOS), as evidenced by pronounced peaks in Figure~\ref{fig3}a,c. These peaks reflect localized electronic states and underscore the unique topological features of Mo$_{2}$ScN$_{2}$O$_{2}$. Figure~\ref{fig3}c depicts the band structure of Mo$_{2}$ScN$_{2}$O$_{2}$ as projected onto the atomic orbitals of Mo and Sc atoms, which include s, d$_{{xy}}$, d$_{{yz}}$,d$_{{xz}}$, d$_{{z}^2}$, and d$_{{x}^2}$$_{{-y}^2}$. It is important to note that the d orbitals of both Mo and Sc are the primary contributors to the electronic states close to the Fermi level, while the p orbitals of N and O atoms have a lesser contribution in this energy range. The saddle-type Van Hove singularity is predominantly influenced by the d$_{{z}^2}$ orbital of Mo and the d$_{{x}^2}$$_{{-y}^2}$ orbital of Sc. To further investigate these singularities, a detailed 3D visualization of the band structure was calculated. Figures~\ref{fig3}a and \ref{fig4} illustrate the three-dimensional profiles of Bands 1 and 2, respectively, allowing for the clear identification of the saddle points that contribute to the Van Hove singularity. The saddle-shaped surface is mathematically represented by the Cartesian equation $z \propto x^{2}-y^{2}$, which corresponds closely with the characteristics seen in the three-dimensional energy band diagrams. This saddle-like topology signifies a Van Hove singularity and is crucial to understanding the electronic properties of Mo$_{2}$ScN$_{2}$O$_{2}$. In the subsequent analysis, we examined the electron-phonon coupling (EPC) associated with Bands 1 and 2, which will be discussed in detail in later sections.

As previously discussed, Mo$_{2}$ScN$_{2}$O$_{2}$ exhibits the highest predicted T$_{c}$ among the o-MXenes examined. One of the most striking features contributing to this behavior is the emergence of a flat electronic band at the Fermi level, which may play a pivotal role in enhancing superconductivity. Band-structure calculations, as illustrated in Figure~\ref{fig3}a,c, reveal a pronounced flat-band segment at the Fermi energy. The presence of this flat band significantly influences two key superconducting parameters: the electron-phonon coupling constant ($\lambda$) and the coherence length ($\xi$), which characterizes the spatial extent of a Cooper pair. These effects can be interpreted in relation to changes in Fermi velocity (v$_{F}$), as flattening of the band typically corresponds to a reduction in v$_{F}$. Since all relevant parameters-$\lambda$, $\xi$, and v$_{F}$-are derived from the presented band and phonon data, a comprehensive understanding of the flat-band influence is accessible within the current framework.   For a superconductor within the framework of conventional BCS theory, the zero-temperature coherence length is given by the following well-known relation\cite{fetter2012quantum}:

\begin{equation}
	\xi(0) = \sqrt{\frac{7 \xi(3)}{3}} \, \frac{\hbar v_F}{K_B 4\pi T_C}
\end{equation}

Table 3S summarizes the coherence lengths ($\xi$(0)) calculated for all stable o-MXenes. Notably, Mo$_{2}$ScN$_{2}$O$_{2}$ exhibits an exceptionally short coherence length of approximately 2 nm, which is markedly smaller than those observed in conventional MXenes \cite{D2NR01939F}. This value is comparable to those reported for well-known superconductors such as hydrides\cite{ABDSHUKOR2021104219}, metal diborides (e.g., MgB$_{2}$)\cite{larbalestier2001strongly}, and high-T$_{c}$ cuprate superconductors\cite{RevModPhys.66.1125}. Furthermore, the presence of a flat band between the M and K points in  Mo$_{2}$ScN$_{2}$O$_{2}$ leads to the emergence of heavy fermionic states near the Fermi level. This feature significantly enhances the electronic density of states (DOS) at the Fermi level, a critical factor known to promote strong electron-phonon coupling (EPC) and support high-T$_{c}$ superconductivity.

Figure 6S illustrates the bonding and antibonding states resolved by orbitals, offering a deeper understanding of the electronic structure of the material. In particular, Figure~\ref{fig3}d shows the crystal orbital Hamilton population (COHP) for Mo$_{2}$ScN$_{2}$O$_{2}$, indicating that the anti-bonding states are mainly filled at the Fermi level. These antibonding electrons, which are less spatially restricted and more delocalized, are ideally situated to engage in Cooper pairing, thus aiding in the increase of the superconducting transition temperature (T$_{c}$). Furthermore, the occupation of antibonding states at the Fermi level leads to an elevated density of states (DOS) in close proximity to the Fermi energy, which significantly amplifies electron correlations and strengthens the overall electron-phonon interaction—key factors in the emergence of superconductivity. 

It is noteworthy that the occupation of antibonding states plays a critical role in driving structural instability in nitride-based o-MXenes. As depicted in Figure~\ref{fig2}, these materials exhibit heightened instability, attributed to the higher valence electron count of nitrogen. The excess electrons partially fill both bonding and antibonding states, weakening interatomic bonds. Specifically, all Cl- and H-functionalized nitride o-MXenes display pronounced structural instability. This instability arises not only from electron occupation of antibonding orbitals but also from surface-induced charge redistribution. To explore this effect, depletion charge analysis was performed, revealing that the surface functional groups accumulate excess charge. This leads to significant electrostatic repulsion between adjacent functionalized atoms, further destabilizing the o-MXene lattice. This charge transfer process in compound W$_{2}$VN$_{2}$T$_{2}$, which is functionalized with atoms O, F, Cl, and H is illustrated in Fig.~7S.

Furthermore, as shown in Figure~\ref{fig2}, carbide-based o-MXenes functionalized with H and Cl exhibit significantly higher T$_{c}$ compared to those functionalized with O and F. This trend is clearly demonstrated in our calculations, where W$_{2}$VC$_{2}$H$_{2}$ (T$_{c}$ = 16.8 K) displays much stronger superconducting behavior than its oxygen-functionalized counterpart W$_{2}$VC$_{2}$O$_{2}$ (T$_{c}$ = 0.3 K). A similar contrast is observed between Mo$_{2}$ScC$_{2}$H$_{2}$ and Mo$_{2}$ScC$_{2}$O$_{2}$, despite both compounds sharing identical M, M$^\prime$, and X elemental compositions. This difference in T$_{c}$ highlights the pivotal role of the surface functional groups (T) in modulating electron-phonon coupling, a key driver of superconductivity in these materials. The chemical nature and electronegativity of the functionalizing atoms likely alter local bonding environments and lattice vibrations, thereby enhancing or suppressing superconducting behavior.  To evaluate the influence of surface termination groups (T) on superconducting properties, three key parameters were examined. 

(I) Variations in the density of states (DOS): as shown in Figure 4S, the contributions of constituent elements to the DOS at the Fermi level vary significantly depending on the functionalization. When O, F, and Cl groups are adsorbed onto M$_{2}$M$^\prime$X$_{2}$ surfaces, each F and Cl atom acquires one electron, while O atoms gain two electrons from the parent structure. This electron transfer lowers the Fermi energy. Conversely, adsorption of H atoms leads to electron loss, shifting the Fermi energy upward. These shifts in the Fermi level alter the Fermi surface topology. According to Figure 5S, functionalization with O, F, and Cl results in Fermi surfaces featuring more band crossings and increased spatial extension. This extended Fermi surface is conducive to the formation of Cooper pairs, as predicted by Bardeen-Cooper-Schrieffer (BCS) theory, and may therefore enhance superconducting performance. In addition to changes in the density of states and Fermi surface topology, two further parameters critically influence the superconducting properties of o-MXenes as discussed below.
 
(II) Phonon dispersion modulation: as mentioned previously, adsorption of O, F, and Cl atoms onto M$_{2}$M$^\prime$X$_{2}$ surfaces leads to electron gain from neighboring metal atoms (M), strengthening the surface-M bonds. In contrast, H atoms exhibit electron loss, resulting in weaker bonding interactions. These variations affect the lattice constant, with H-functionalized structures typically exhibiting an expansion. This weaker bonding induces phonon softening, particularly in the acoustic modes, as evident in Figure 2S. These soft phonon modes play a crucial role in enhancing the electron-phonon coupling (EPC) and T$_{c}$. 

(III) Electron-phonon coupling parameters: Further evidence from the Eliashberg spectral function $\alpha$$^{2}$F($\omega$) in Figure 9S indicates a pronounced increase in the electron-phonon coupling constant ($\lambda$) for select functionalizations. This rise in $\lambda$ correlates strongly with the observed enhancement in T$_{c}$, confirming the significant contribution of surface terminations in modulating EPC strength. These mechanisms are thoroughly investigated in the following sections. 

\subsection{Electron–Phonon Coupling and Superconductivity}
To deepen our understanding of superconductivity in o-MXenes, we analyzed their phonon dispersion and electron-phonon coupling (EPC) behavior. As shown in Figures~\ref{fig:ph} and 2S, these materials exhibit robust dynamical stability, as evidenced by the absence of imaginary phonon frequencies throughout the Brillouin zone. The phonon spectra divide into two distinct regions due to atomic mass differences. The low-frequency regime (0-42 meV) is dominated by vibrational modes involving heavier transition metals (M = Mo, W, and M$^\prime$ = Sc, Y, Ti, etc.). The high-frequency regime ($>$42~meV), governed by lighter functional groups (Cl, F, O, H) and C/N atoms, features elevated vibrational frequencies that contribute less effectively to EPC (see Figure~\ref{fig:ph}a). As illustrated in Figures~\ref{fig:ph}a and 2S, phonon linewidths ($\gamma_{q\nu}$) are generally larger in the high-frequency region. However, since the EPC constant ($\lambda$) scales inversely with the square of the phonon frequency ($\lambda$ $\propto$ $\gamma_{q\nu}$/$\omega^{2}$), low-frequency M/M$^\prime$ vibrations—despite their smaller $\gamma_{q\nu}$ play a pivotal role in amplifying $\lambda$. According to calculations of the Eliashberg spectral function, $\alpha^{2}$F($\omega$), approximately 65$\%$ of the total EPC coefficient ($\lambda$) originates from phonon modes with frequencies below 42~meV. Figure 9S presents $\alpha^{2}$F($\omega$) spectra for all stable o-MXenes, demonstrating this trend across the material family. Since electron-phonon matrix elements are band-index dependent, we next examine the specific role of Fermi-level band crossings.

In Mo$_{2}$ScN$_{2}$O$_{2}$, as discussed previously, the electron-phonon coupling primarily originates from electrons in the flat band (first band) and the second band intersecting the Fermi level. To investigate this further, we analyze the EPC matrix within the flat band using the QuantumATK V-2016.4 \cite{smidstrup2019quantumatk} package. This approach calculates the transition rate due to phonon scattering from an electronic state $\ket{kn}$ to $\ket{k'n'}$. Comparing low-frequency and high-frequency phonon modes reveals distinct behaviors, attributed to the intrinsic nature of these phonons and their coupling strength to electronic states.
Low-frequency modes—typically associated with heavier atomic vibrations—show stronger coupling characteristics in the context of EPC enhancement, while high-frequency modes—often involving lighter atoms—exhibit comparatively weaker coupling to the relevant electronic states.  Figures 8S depict the electron-phonon coupling (EPC) matrix elements derived from Equation S14, shown as functions of various physical parameters: the phonon wavevector q (near the Brillouin zone center), the phonon dispersion of the first acoustic mode, the electronic states before and after phonon absorption, and, ultimately, the phase space for phonon absorption and emission processes.

In Mo$_{2}$ScN$_{2}$O$_{2}$, the electronic states involved in EPC—both the initial and final states during electron-phonon scatterings—are confined to the first and second conduction bands. The electronic wavevector k is fixed at the M point of the Brillouin zone for these calculations. We investigated EPC contributions from the first five phonon bands, which include three acoustic and two lowest optical bands. This mode-resolved analysis allows us to pinpoint the dominant scattering pathways and assess the comparative impact of vibrational character on EPC strength.  Figure~\ref{fig4} presents the three-dimensional band structure alongside a cross-sectional view of the Fermi surface and the corresponding electron-phonon coupling (EPC). The first acoustic phonon mode interacts significantly with the first conduction band, resulting in a pronounced EPC contribution. In contrast, subsequent acoustic modes possess higher phonon energies and display weaker EPC. Figure 8S reveals an inverse relationship between EPC strength and phonon energy $\hbar\omega_{q}$. In regions of the Brillouin zone where phonon energies are lower, EPC tends to be stronger. This trend is further reinforced in higher-order phonon bands, where elevated EPC contributions emerge from energetically favorable scattering conditions. For phonon energies $\hbar\omega_{q} $ $<$ 0.01 eV, certain electronic transitions across nearby q-points satisfy the delta-function condition, as indicated by the yellow regions in the $\delta$-function maps of Figure 8S. These regions correspond to strong scattering channels. Conversely, high-frequency optical modes—such as phonon mode 21—exhibit relatively flat and elevated dispersion profiles (Figure~\ref{fig4}). These optical phonons, characterized by out-of-phase atomic vibrations within the unit cell, couple less efficiently with electronic states near the Fermi energy, resulting in reduced EPC contributions.

 The second conduction band in Mo$_{2}$ScN$_{2}$O$_{2}$ features a steeper energy slope, indicating higher electron velocity, increased kinetic energy, and reduced spatial localization. As illustrated in Figure~\ref{fig4}, the electrons in this band possess approximately twice the energy of those in the first band. Despite the availability of suitable phonon modes, these electrons are unable to satisfy the $\delta$-function condition essential for strong EPC interactions, thereby impeding the formation of Cooper pairs. In high-frequency phonon modes, the elevated phonon energy acts as a constraint on electron mobility and limits the scattering phase space, leading to only marginal EPC formation. This behavior is supported by multiple visual representations, e.g., Figure 8S.d,h shows high phonon energies at the Brillouin zone center; the $\delta$-function map in Figure 8S.d,h identifies regions lacking adequate overlap between electronic states and phonons; and the EPC intensity distribution confirms the diminished interaction strength. Therefore, in the second band, EPC strength and phonon energy $\hbar\omega_{q}$ tend to follow a direct correlation, unlike the inverse behavior observed in lower-energy bands. The EPC map for each band reveals an anisotropic distribution, indicating that this effect must be accounted for in calculations. 

Advancing beyond the harmonic and isotropic framework, our analysis now turns to two phenomena that are critical for an accurate microscopic model. In the following section, we will check 2 phenomena that have a major effect on superconductivity. First, Anharmonicity in the phonon spectrum can significantly alter both phonon lifetimes and electron–phonon coupling strengths, leading to deviations from predictions based on the harmonic approximation. In many systems, anharmonicity softens specific phonon modes, thereby enhancing the pairing interaction, while in others it introduces additional scattering channels that suppress superconductivity~\cite{Mishra2025-yy}\cite{Mishra2026}. Secondly, Anisotropy in the electronic structure and electron–phonon interaction also plays a crucial role. While isotropic models assume uniform pairing over the Fermi surface, real materials often exhibit anisotropic gaps due to band-structure complexity or directional variations in the coupling strength~\cite{margine2013anisotropic,Mishra2024}. Accounting for this anisotropy generally leads to more accurate predictions of the superconducting gap distribution and critical temperature, particularly in materials with multiple Fermi-surface sheets or low-symmetry crystal structures. 

To comprehensively assess T$_{c}$, therefore we employ a suite of six theoretical methods and computational frameworks, a wide range. This variation arises naturally from the distinct levels of approximation, underlying physical assumptions, and numerical implementations inherent to each approach. Semi-empirical models such as the McMillan and Allen–Dynes equations rely on simplified parameterizations and tend to provide lower-bound or approximate estimates of T$_{c}$. 
In contrast, fully self-consistent solutions of the isotropic and anisotropic Eliashberg equations capture the detailed electron–phonon interaction spectrum and yield more rigorous predictions.

\subsection{Anharmonic and Anisotropic Effects on Superconductivity}

\begin{figure*}
\includegraphics[width=1.0\linewidth]{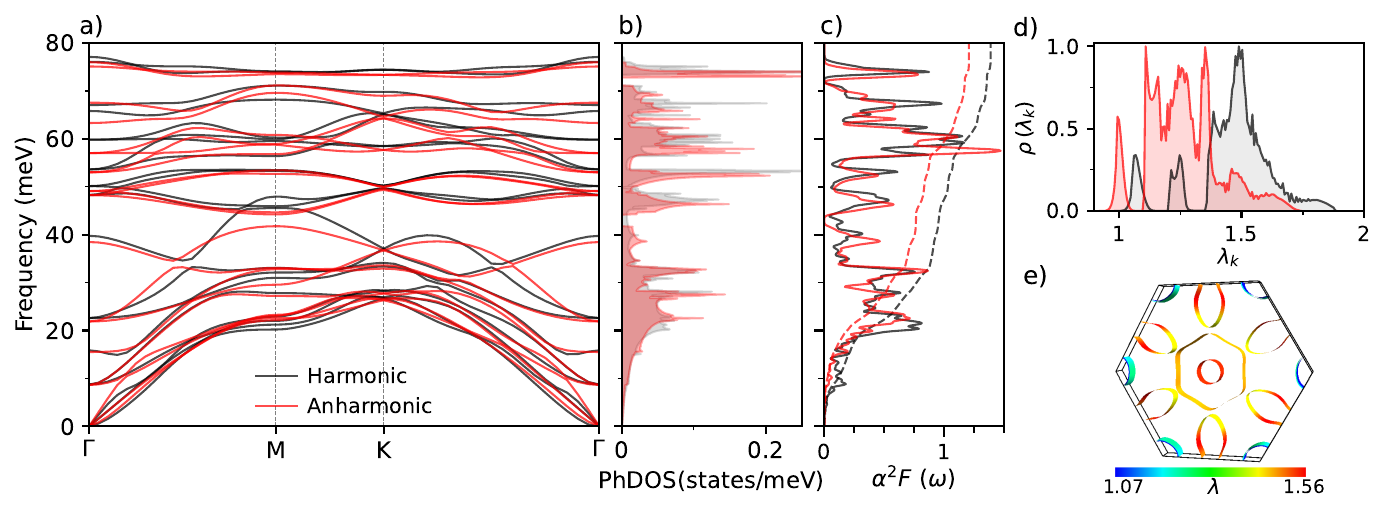}
\caption{\label{fig:wide}a) Phonon dispersion, b) total phonon density of states (PhDOS), c) isotropic Eliashberg spectral function $\alpha^2F(\omega)$ (solid line) and cumulative electron-phonon coupling strength $\lambda(\omega)$ (dashed lines), and d) distribution of the electron-phonon coupling strength $\lambda_{\textbf{k}}$ for both harmonic (black) and anharmonic (red) phonons in  Mo$_{2}$ScN$_{2}$O$_{2}$. (e) Momentum-resolved electron-phonon coupling $\lambda_{\textbf{k}}$ mapped onto the Fermi surface (shown for the harmonic case).}\label{fig:ph}
\end{figure*}

The roles of anharmonicity and anisotropy in determining superconducting properties have been examined for our high-T$_{c}$ candidate, Mo$_{2}$ScN$_{2}$O$_{2}$. Anharmonic phonon corrections are negligible for the heavier atomic modes, which correspond to the low-frequency Mo and Sc modes below 42~meV.
In our calculations, the SSCHA minimization was performed under a constant-volume relaxation scheme, which introduces a quantum pressure that further contributes to the anharmonic landscape\cite{D5MH00177C}\cite{PhysRevB.106.134509}:

\begin{equation}
p_q = -\frac{1}{\Omega} \frac{\partial f}{\partial \varepsilon} \bigg|_{\varepsilon = 0}.
\end{equation}
Here, $p_q$ originates from the quantum zero-point energy (ZPE) and thermal expansion at 300 K, with $\Omega$, $f$, and $\varepsilon$ denoting the cell volume, SSCHA free energy, and harmonic lattice strain, respectively. Incorporating these effects leads to an expansion of the lattice constant, which in turn softens phonon modes and reduces the EPC constant ($\lambda$).

Figure~\ref{fig:ph}a-b compares the phonon dispersion and phonon density of states (PhDOS) for Mo$_{2}$ScN$_{2}$O$_{2}$, calculated within harmonic and anharmonic approximations. The overall spectral features remain similar, however, anharmonic corrections induce a slight softening of the phonon modes. Within the harmonic approximation, the total EPC strength is $\lambda = 1.39$. The low-frequency phonon modes below 40~meV, associated primarily with vibrations of the heavier Mo and Sc atoms, contribute about 67\% to $\lambda$, while the remaining 33\% originates from high-frequency vibrations of the lighter N and O atoms. Incorporation of anharmonic effects reduces the total EPC strength by $\sim$13\%, yielding $\lambda = 1.21$. In this case, the contribution of low-frequency modes decreases to 62\%, whereas the relative weight of high-frequency modes increases to 38\%. 

\begin{figure*}
\includegraphics[width=1.0\linewidth]{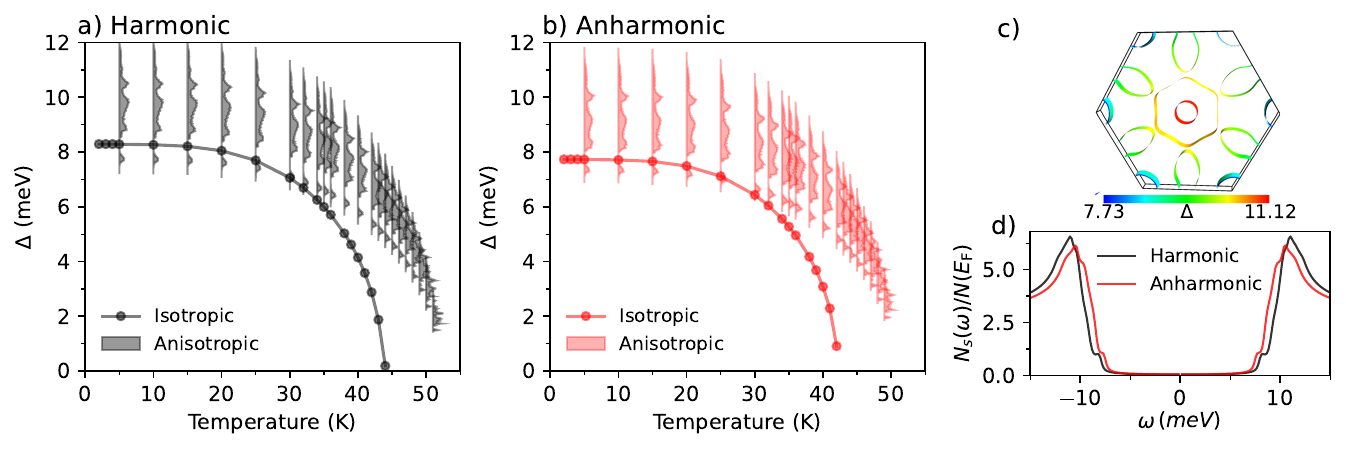}
\caption{\label{fig:wide}Temperature dependence of the superconducting gap $\Delta$ for Mo$_{2}$ScN$_{2}$O$_{2}$, showing isotropic and anisotropic solutions for a) harmonic and b) anharmonic phonons. c) Momentum-resolved superconducting gap $\Delta{\mathbf{_{k}}}$ on the Fermi surface, visualized with FermiSurfer~\cite{Kawamura2019}. d) Normalized quasiparticle density of states (DOS) calculated at $T=5$~K for $\mu^*=0.1$, comparing harmonic and anharmonic cases.}
\label{fig:gap}
\end{figure*}

\begin{table*}[t]
\centering
\caption{Properties of Mo$_{2}$ScN$_{2}$O$_{2}$: the EPC $\lambda$, logarithmic average phonon frequency $\omega_{\rm log}$, and the superconducting critical temperature $T_{\rm c}$ and zero-temperature gap $\Delta_0$. Critical temperatures are compared using the Allen--Dynes (AD) formula, and the isotropic and anisotropic FBW approaches for both harmonic and anharmonic phonons with a Coulomb pseudopotential $\mu^* = 0.1$.}
\label{tab:tc}

\begin{tabular}{c|c c c c c c c c}
phonons &
$\lambda$ &
$\omega_{\rm log}$ (meV) &
$T_{\rm c}^{\rm AD}$ (K) &
$\Delta_0^{\rm iso}$ (meV) &
$T_{\rm c}^{\rm iso}$ (K) &
$\Delta_0^{\rm aniso}$ (meV) &
$T_{\rm c}^{\rm aniso}$ (K) \\
\hline\hline
Harmonic   & 1.39 & 27.8 & 38.4 & 8.2 & 44 & 10 & 52 \\
Anharmonic & 1.21 & 32.7 & 37.4 & 7.8 & 42 & 10 & 48 \\
\hline\hline
\end{tabular}
\end{table*}

We further investigated the effect of anisotropy by computing the momentum-resolved coupling $\lambda_{\mathbf{k}}$ as defined in Ref.~\cite{margine2013anisotropic}. In the harmonic case Fig.~\ref{fig:ph}(c), the distribution displays two narrow peaks and a broader peak with a maximum centered around $\lambda_{\mathbf{k}} \approx 1.5$. Upon inclusion of anharmonic effects, the distribution shifts toward lower values, yielding a smaller peak near 1.0 and a broader feature spanning 1.1–1.5. The momentum-resolved distribution of $\lambda_{\mathbf{k}}$ on the Fermi surface (FS), shown in Fig.~\ref{fig:ph}(d), provides further insight into the origin of these features. The first, low-magnitude peak is associated with the hexagonal corner regions of the FS, primarily derived from Sc and O states (see Fig.~\ref{fig3}), reflecting their nearly free-electron-like character in the band structure. The broader second peak corresponds to the red regions of the FS and originates from contributions of all states.

To determine the superconducting properties, we solved both the isotropic and anisotropic Migdal-Eliashberg equations~\cite{margine2013anisotropic}. Given the presence of a VHS peak close to the Fermi level (see Fig.\ref{fig3}), we employed the full-bandwidth (FBW) approach, which incorporates the full energy dependence of the DOS and accounts for electron–phonon scattering processes away from the Fermi level~\cite{Lee2023,Lucrezi2024}. For the anisotropic calculations, the sparse-IR sampling technique was used to efficiently reduce the number of Matsubara frequency sampling points without loss of accuracy~\cite{Wallerberger2023,Mori2024}. 

Figure~\ref{fig:gap}(a,b) shows the temperature dependence of the isotropic and anisotropic superconducting gaps, computed within the FBW framework using a Coulomb pseudopotential $\mu^* = 0.1$, for both harmonic and anharmonic phonons. The isotropic and anisotropic solutions yield critical temperatures $T_{\rm c}$ of 44 and 52 K, respectively, with corresponding zero-temperature gaps of $\Delta_0 \approx 8.2$ and $\sim$10 meV. Inclusion of anharmonic effects reduces $T_{\rm c}$ by approximately 4~K, as summarized in Table~\ref{tab:tc}. The anisotropic calculations consistently predict slightly higher $T_{\rm c}$ values compared to the isotropic ones. The results indicate a continuous anisotropic gap without splitting into distinct gaps, similar to what has been reported in CaC$_6$~\cite{Margine2016}.

To further characterize the gap structure, we plot the momentum-resolved superconducting gap $\Delta_{\mathbf{k}}$ on the Fermi surface at 5 K for in the harmonic case. The lower peak at $\Delta_1 \approx 8.2$ meV originates from the Brillouin-zone corner regions, consistent with the $\lambda_{\mathbf{k}}$ distribution, while the broader peak corresponds to states near the zone center and side regions. This continuous anisotropic gap structure is further supported by the normalized quasiparticle DOS, shown in Fig.~\ref{fig:gap}(d).

\section{Computational Details}
All structural optimizations, electronic structure analyses, and ab initio molecular dynamics (AIMD) simulations were conducted using density functional theory (DFT) within the Quantum ESPRESSO package\cite{giannozzi2009quantum}\cite{giannozzi2017advanced}\cite{giannozzi2020quantum}. The exchange–correlation functional was approximated via the Perdew-Burke-Ernzerhof (PBE) generalized gradient approximation (GGA)\cite{hamann2013optimized}\cite{schlipf2015optimization}. Ultrasoft pseudopotentials were employed with kinetic energy and charge density cutoffs set to 70 Ry and 700 Ry, respectively. The total energy was achieved with convergence thresholds of $10^{-8}$ Ry. Structural relaxation was carried out using the BFGS algorithm until the maximum force acting on any atom fell below $10^{-7}$ Ry/Bohr. Density functional perturbation theory (DFPT)\cite{baroni2001phonons} calculations were executed to derive phonon dispersion relations and electron-phonon coupling (EPC) parameters within the linear-response framework. A 12 $\times$ 12 $\times$ 1 K-point mesh was used for self-consistent field (SCF) calculations, while a finer 48 $\times$ 48 $\times$ 1 K-point grid ensured accuracy in EPC matrix evaluations. Phonon properties were computed using a 6 $\times$ 6 $\times$ 1 q-point grid.
Material stability was assessed via formation energy calculations (Equation S5, Supplementary Information). Bonding interactions were quantified through crystal orbital Hamilton population (COHP)\cite{doi:10.1021/j100135a014} analysis using the LOBSTER code, which decomposes bonding contributions into antibonding, nonbonding, and bonding states. The integrated COHP (ICOHP) up to the Fermi level served as a metric for bond strength. Fermi velocities ($v_{F}$) were derived from band structures using $v_{F} = \hbar^{-1} \nabla_{\mathbf{k}} \varepsilon_{\mathbf{k}} \bigg|_{\varepsilon = E_{F}}$. 

The Eliashberg spectral function and McMillan equation were applied to estimate the T$_{c}$ and isotropic EPC strength. For anisotropic effects, the full Eliashberg equations\cite{margine2013anisotropic} were utilized. Also, we checked the electron-phonon coupling distribution in our highest T$_{c}$ candidate using QuantumATK 2016.4 \cite{smidstrup2019quantumatk}\cite{PhysRevB.93.035414}\cite{PhysRevB.85.115317}\cite{PhysRevLett.109.166604} software.  The SSCHA\cite{monacelli2021stochastic} and EPW\cite{giustino2007electron}\cite{ponce2016epw} codes were employed to account for anharmonic and anisotropic effects on the superconducting properties of Mo$_{2}$ScN$_{2}$O$_{2}$.

\section{Conclusion}
In brief, our first-principles search shows that surface-functionalized o-MXenes are a promising class of two-dimensional superconductors with tunable properties. Out of 128 candidates, 42 compounds are mechanically, dynamically, and thermodynamically stable with superconducting transition temperatures ranging from 0.1 K to 52 K. Mo$_{2}$ScN$_{2}$O$_{2}$ stands out as the most promising candidate, exhibiting the highest predicted transition temperature (T$_{c}$ $\approx$ 52 K) among MXenes. This high Tc originates from an anisotropic superconducting state underpinned by flat-band effects and contributions from anharmonic lattice dynamics. The results emphasize the significant roles played by surface functionalization, orbital hybridization, and anisotropy in governing electron–phonon coupling and superconductivity. Besides identifying Mo$_{2}$ScN$_{2}$O$_{2}$ as a potential high-T$_{c}$ material, this study provides design ideas that can serve to guide the targeted synthesis of new MXene-based superconductors in addressing the general pursuit of future two-dimensional superconducting materials.

\medskip

\medskip
\textbf{Acknowledgements} 

S.B.M and E.R.M. acknowledge the National Science Foundation (NSF), Office of Advanced Cyberinfrastructure under Grant No. 2103991. Anisotropic superconductivity calculations were carried out using Stampede3 supercomputers at the Texas Advanced Computing Center (TACC) at the University of Texas at Austin, supported through the ACCESS allocation TG-DMR180071.HR was funded by JSPS KAKENHI (No. 23K04361). K.M. was supported by JSPS KAKENHI Grant Numbers 25K07220, 24H00190.

\bibliography{apssamp}

\medskip
\textbf{Author contributions} \newline
M.K. conceptualization, methodology, calculations, formal analysis, investigation, writing original draft. F.D. performed part of the calculations and reviewed the manuscript. S.B.M and E.M. methodology, performed part of the calculations, writing, reviewing, and editing. M.S, M.J.A., M.Ka., and H.R. writing, reviewing, and editing. R.P.P, K.H, R.M, and M.H.M writing, review, and editing, computational facilities. M.Kh. project administration, designed and supervised the project, writing, reviewing, and editing. All authors have read and approved the final version of the manuscript.\newline

\medskip
\textbf{Competing interests} \newline
The authors declare no competing interests.

\end{document}